\documentclass[aps,prl,reprint,superscriptaddress,floatfix]{revtex4-2}
\usepackage{amsmath,amssymb,amsfonts,bm,mathtools}
\usepackage{graphicx}
\usepackage{xcolor}
\usepackage{tikz}
\usetikzlibrary{arrows.meta,calc,positioning}
\usepackage[colorlinks=true,linkcolor=blue,citecolor=blue,urlcolor=blue]{hyperref}
\usepackage{booktabs}
\usepackage{microtype}

\newcommand{\Tr}{\operatorname{Tr}}

\newcommand{\dd}{\mathrm d}

\newcommand{\Om}{\Omega_g}
\newcommand{\bk}{\mathbf k}
\newcommand{\br}{\mathbf r}
\newcommand{\calL}{\mathcal L}
\newcommand{\psiel}{\psi_{\rm el}}
\newcommand{\prlsection}[1]{\paragraph*{#1---}}
\definecolor{softgreen}{RGB}{218,238,224}
\definecolor{softpurple}{RGB}{226,220,239}
\definecolor{softorange}{RGB}{245,229,211}
\definecolor{softblue}{RGB}{65,105,180}
\definecolor{linegray}{RGB}{70,70,70}

\begin{document}

\title{Topological superconductivity from Abelian fractional Chern insulators}
\author{Taige Wang}
\affiliation{Department of Physics, Harvard University, Cambridge, Massachusetts 02138, USA \looseness=-2}
\affiliation{Materials Research Laboratory, Massachusetts Institute of Technology, Cambridge, MA 02139, USA \looseness=-2}
\date{\today}

\begin{abstract}
Can a Laughlin anyon fluid become a topological superconductor? We answer this question for a \(\nu=1/3\) fractional Chern insulator (FCI) using a \(U(3)\) infrared parton theory. Three charge-\(e/3\) constituents form the electron, while three constituent pairs form a gauge-invariant charge-\(2e\) Cooper pair. The resulting superconductors share an ordinary charge-\(2e\) sector but differ in their neutral color response, giving an SC\(^\ast\) with the parent Laughlin \(U(1)_3\) order, a chiral topological superconductor with \(c_-=3/2\), and a strong-pairing molecular superconductor with \(c_-=3\). The same framework organizes a nearby \(\sigma_{xy}=0\) charge density wave (CDW) state. At commensurate filling, the normal state and the \(c_-=3\) and \(c_-=3/2\) superconducting descendants are naturally tied to a period-three density order background, while the SC\(^\ast\) branch can preserve microscopic translations. Away from commensuration, a chargon metal can pair into the same \(c_-=3/2\) topological superconductor. The FCI, reentrant CDW, and chiral superconductivity are unified in the \(U(3)\) parton theory.
\end{abstract}

\maketitle

\prlsection{Introduction}
Anyons carry fractional charge and fractional statistics~\cite{LeinaasMyrheim1977,Wilczek1982,ArovasSchriefferWilczek1984}.  Soon after their introduction, Laughlin and Chern-Simons approaches suggested that a finite-density gas of charged anyons could superconduct~\cite{Laughlin1988PRL,Laughlin1988Science,FetterHannaLaughlin1989,ChenWilczekWittenHalperin1989,LeeFisher1989,Halperin1992}.  Fractional Chern insulators (FCIs) sharpen this question because their anyons live on a lattice, carry crystal quantum numbers, and can acquire dispersive bands~\cite{TangMeiWen2011,SunGuKatsuraDasSarma2011,NeupertSantosChamonMudry2011,ShengGuSunSheng2011,RegnaultBernevig2011}. This problem is now concrete in moir\'e Chern bands.  Fractional anomalous Hall and FCI regimes have been observed in twisted transition-metal dichalcogenides~\cite{CaiFQAH2023,XuFQAH2023,ParkFQAH2023} and rhombohedral graphene moir\'e superlattices~\cite{LuGrapheneFQAH2024,WatersPentalayerGraphene2025}.  Superconductivity has recently been reported nearby in both twisted MoTe$_2$ and rhombohedral graphene platforms~\cite{XuMoTe2SC2025,HanChiralGraphene2025,ChoiGrapheneSCQAH2025}.

A key numerical motivation is the chiral $f$-wave superconducting phase found near a spin-polarized FCI at both $\nu=2/3$ and finite doping~\cite{WangZaletel2025}.  After particle-hole transformation, this phase is a charge-$2e$ topological superconductor with $c_-=3/2$ at $\nu = 1/3$.  Related flat-Chern-band calculations also find chiral $f$-wave superconductivity near fractionalized Chern-band regimes~\cite{GuerciAbouelkomsanFu2025,GuerciAbouelkomsanFu2026VortexLattice}.  The same experimental and numerical phenomenology also contains reentrant insulating or Hall features.  In the orientation used below, the relevant nearby normal descendant is a $\sigma_{xy}=0$ charge density wave (CDW) state.  We ask whether an Abelian Laughlin-type FCI can organize the FCI, this normal descendant, and the $c_-=3/2$ superconductor within one framework.

Several Abelian FCI descendants are known.  Doping or bandwidth tuning can produce topologically trivial anyon superconductors~\cite{ShiSenthil2025PRX,WangZaletel2025,NosovHanKhalaf2026,DivicEtAl2025,PichlerEtAl2026,GaoWangYangWu2025Charge2ne,GaoWangZhangYang2026Charge4e}.  Topological anyon superconductors with an odd number of Majorana edge modes are more constrained.  Existing controlled QCD$_3$--Chern-Simons constructions produce topological superconducting descendants of Abelian FCIs, but do not directly realize the charge-$2e$, $c_-=3/2$ topological superconductor indicated above~\cite{MaHe2020,ShiSenthil2025PRX}.

We give such a route.  The guiding analogy is superconductivity from a fractionalized spin liquid, where chargons condense and spinons pair~\cite{Anderson1987,BaskaranZouAnderson1987,KotliarLiu1988,SenthilFisher2000,LeeNagaosaWen2006,WenPSG2002}.  Here the spinons are replaced by $U(3)$ composite fermions which represent charge-$e/3$ Laughlin anyons.  The formalism keeps track of charge, flux quantization, and pair Higgsing: one electron fractionalizes into three Abelian anyons, and superconductivity arises when three anyon pairs combine into an ordinary charge-$2e$ Cooper pair.

Starting from a $\nu=1/3$ FCI, we obtain charge-$2e$ anyon superconductors with a common charge sector.  In the color-isotropic branch, the neutral descendants are an SC$^\ast$ with residual Laughlin \(U(1)_3\) topological order, a strong-pairing molecular superconductor with $c_-=3$, and an $SO(3)^s_1$ topological superconductor with chiral $f$-wave electron pairing, three chiral Majorana edge modes, and $c_-=3/2$.  At finite density, doped anyons form a chargon metal~\cite{ZhangHolonMetal2025} whose weak pairing in the target chirality reaches the same $SO(3)^s_1$ infrared superconducting response.  Zero doping and finite doping are therefore unified in the same determinant-locked $U(3)$ Higgs theory.

We also find a \(\sigma_{xy}=0\), \(c_-=0\) Mott descendant.  In a microscopic Hofstadter realization, the \(2\pi/3\) magnetic-translation algebra forces it into a period-three charge density wave (CDW), matching after particle-hole transformation the RIQH features seen in numerics and experiment~\cite{WangZaletel2025,XuMoTe2SC2025}.  The commensurate \(SO(3)^s_1\) superconductor is tied to the same period-three background and coexists with density order, even though its gauge-invariant Cooper pair has zero momentum.  Experiments cannot yet resolve whether the superconducting phase breaks translations~\cite{XuMoTe2SC2025,HanChiralGraphene2025}, but numerics show suggestive hints~\cite{WangZaletel2025}.

Our main construction assumes translation breaking, which lifts the magnetic-valley multiplet and leaves one low-energy sector per color.  We also describe a translation-symmetric construction in which a full-rank color-valley pair field Higgses the neutral $SU(3)$ gauge group and gives a translation-preserving charge-$2e$, $f-if$, $c_-=3/2$ superconductor~\cite{FanVishwanathWang2026}.

\prlsection{$U(3)$ partons and magnetic translations}
Before entering the formalism, we outline the intuition behind the construction.  In a Laughlin $\nu=1/3$ state, an electron fractionalizes into three charge-$e/3$ constituents,
\begin{equation}
   \psiel\sim f_1f_2f_3 .
   \label{eq:schematicparton}
\end{equation}
The $f_a$ are composite fermions, not physical anyon-creation operators.  In the Laughlin branch below, a dressed $f$ line carries charge $e/3$, has the Laughlin quasiparticle statistics, and can therefore be identified with the charge-$e/3$ anyon.  Most anyon-superconductor constructions put the three constituents in different states~\cite{ShiSenthil2025PRX,WangZaletel2025,NosovHanKhalaf2026,DivicEtAl2025,PichlerEtAl2026,GaoWangYangWu2025Charge2ne,GaoWangZhangYang2026Charge4e}. Here we keep them equivalent by promoting $a$ to $SU(3)$ gauge color.  Within this composite fermion picture, a $C_f=1$ band gives the familiar $\nu=1/3$ FCI~\cite{Wen1999Projective}, a $C_f=0$ band gives the confined $\sigma_{xy}=0$ Mott/CDW descendant, and most interestingly pairing the $f$ partons gives the desired topological superconductor. The non-Abelian gauge structure simply enforces this equivalence while tracking the flux lattice and magnetic translations.

We now implement Eq.~\eqref{eq:schematicparton} as a parton construction:
\begin{equation}
   \psiel=b\,f_1f_2f_3
   = \frac{b}{3!}\epsilon_{abc}f_af_bf_c .
   \label{eq:parton}
\end{equation}
The second form displays the determinant representation.  The field
$f=(f_1,f_2,f_3)^T$ is a fundamental of
\begin{equation}
   U(3)=\frac{SU(3)\times U(1)}{\mathbb Z_3},
   \label{eq:U3quotient}
\end{equation}
and carries no microscopic electromagnetic charge.  The determinant Higgs
field $b$ transforms as $\det^{-1}$ under $U(3)$ and has physical charge
$+1$.  Writing
\begin{equation}
   c=\alpha+a\mathbf 1_3,
   \qquad a=\frac{1}{3}\Tr c,
   \qquad \Tr\alpha=0 ,
   \label{eq:decompose}
\end{equation}
the covariant derivatives are
\begin{equation}
\begin{aligned}
   D_\mu f&=(\partial_\mu-i c_\mu)f,\\
   D_\mu b&=(\partial_\mu-iA_\mu+i\Tr c_\mu)b .
\end{aligned}
\label{eq:covder}
\end{equation}
Condensing $b$ is invariantly described by
\begin{equation}
   \calL_b=\frac{1}{2\pi}\,\beta_b\,\dd(A-\Tr c),
   \label{eq:detlock}
\end{equation}
with compact dual gauge field $\beta_b$.  Locally this gives $a\simeq A/3$, or equivalently $f$ carries $e/3$ charge~\footnote{In a lifted $SU(3)\times U(1)$ notation, a
charge-$-3$ Higgs might appear to leave a separate $\mathbb Z_3$ gauge
group.  In the quotient group of Eq.~\eqref{eq:U3quotient}, that subgroup
is identified with the $SU(3)$ center.  Determinant locking therefore
leaves the local neutral algebra $su(3)$ while preserving the correct
$U(3)$ flux lattice and line selection rule.}. At this level it is determinant locking,
not superconductivity. After determinant locking, a dressed fundamental $f$ line sees $A/3$ and carries charge $e/3$, so $f_1f_2f_3$ is the electron fractionalized into three Laughlin anyons.

A translation-preserving microscopic description contains three magnetic valleys for each of the three gauge colors,
\begin{equation}
   f_{aI},
   \qquad a=1,2,3,
   \qquad I=1,2,3,
   \label{eq:ninePartons}
\end{equation}
where $a$ is $SU(3)$ gauge color and $I$ is magnetic valley.  For every fixed color, the charge-$e/3$ parton sees $2\pi/3$ flux per microscopic unit cell and obeys
\begin{equation}
   T_1T_2T_1^{-1}T_2^{-1}=\omega,
   \qquad \omega=e^{2\pi i/3}.
   \label{eq:magneticAlgebraMain}
\end{equation}
The minimal projective representation is three dimensional, giving nine slow fields in total.  The equivalent color-rotation and magnetic-unit-cell gauges are given in the End Matter.

Translation breaking lifts the magnetic-valley multiplet and allows one valley sector per color to participate in the normal and BdG inversions.  We relabel the resulting three fields as $f_a$.  The other six microscopic modes remain gapped and do not participate in the inversions considered below.  A translation-symmetric construction is discussed later.

\prlsection{Normal descendants}
Our convention for a $U(3)$ Chern-Simons response is
\begin{equation}
\calL_{U(3)_{k,r}}[c]
=\frac{k}{4\pi}\Tr\left(c\dd c+\frac{2}{3}c^3\right)
+\frac{r-k}{12\pi}(\Tr c)\dd(\Tr c),
\label{eq:U3kl}
\end{equation}
so the trace field $a=(1/3)\Tr c$ has level $3r$.  A $C_f=1$ band of the $U(3)$ fundamental gives the spin theory $U(3)^s_{1,1}$.  After determinant locking, it gives the standard $SU(3)^s_1$ description of the $\nu=1/3$ FCI with $\sigma_{xy}=(1/3)e^2/h$ and $c_-=1$ from the level-rank duality~\cite{Wen1999Projective,MaHe2020,HsinSeiberg2016},
\begin{equation}
   3\Om-SU(3)^s_1\simeq U(1)_3,
   \qquad c_-=1.
   \label{eq:levelrank}
\end{equation}
Here $\Om$ denotes the gravitational Chern-Simons term.  The counterterm $3\Om$ supplies three complex chiral modes, while gauging color removes the $SU(3)_1$ current algebra of central charge two, leaving $c_-=1$.  The superscript $s$ denotes a spin gauge field.

A $C_f=0$ band gives $U(3)^s_{0,0}$.  The neutral $SU(3)^s_0$ gauge theory is expected to confine, leaving $\sigma_{xy}=0$ and $c_-=0$.  In the microscopic Hofstadter description, changing $C_f$ from one to zero requires translation breaking, giving the Mott/CDW descendant associated with the reentrant insulating or Hall feature.  The same conclusion follows from the filling-$1/3$ LSM constraint, which forbids a featureless translation-preserving insulator~\cite{Oshikawa2000,Hastings2004,WatanabePoVishwanath2015}.

\prlsection{Pair Higgs and charge sector}
We now work with these three fields.  The simplest pair Higgs is
\begin{equation}
    P_{ab}=P_{ba}\in {\bf 6}={\rm Sym}^2({\bf 3}),
    \qquad P\mapsto gPg^T .
    \label{eq:Prep}
\end{equation}
A local Hubbard-Stratonovich coupling is
\begin{equation}
   \calL_Y=
   \eta P^\dagger_{ab}f_aD_- f_b+{\rm h.c.},
   \qquad D_-=D_x-iD_y .
   \label{eq:Yukawa}
\end{equation}
Since $P_{ab}=P_{ba}$, Fermi statistics requires odd orbital angular momentum.  We focus on the lowest channel with chirality opposite to the $C_f=1$ anyon band.  For a weakly paired color band,
\begin{equation}
   C_{\rm BdG}=2C_f+\ell .
\end{equation}
The $p_x-ip_y$ channel has $\ell=-1$, giving $C_{\rm BdG}=1$ and one chiral Majorana mode per color.  The opposite chirality has $\ell=+1$ and $C_{\rm BdG}=3$ per color mode, and represents a distinct higher-chirality descendant with a different residual topological order.

The gauge-invariant physical Cooper pair is
\begin{equation}
   \Delta_{2e}=b^2\det P,
   \qquad
   \det P=\frac{1}{6}\epsilon_{abc}\epsilon_{def}P_{ad}P_{be}P_{cf} .
   \label{eq:Cooper}
\end{equation}
Since $\det P\mapsto(\det g)^2\det P$, the operator $b^2\det P$ is gauge invariant and carries physical charge $2e$.  In a color-diagonal gauge, suppressing the odd-parity form factor,
\begin{equation}
   \Delta_{2e}\sim b^2(f_1D_-f_1)(f_2D_-f_2)(f_3D_-f_3),
   \label{eq:sixanyonpair}
\end{equation}
namely a Cooper pair made from six charge-$e/3$ anyons.

Only the trace charges of the two condensates determine the physical charge response.  The determinant Higgs couples to $A-3a$, while the common phase of a full-rank pair Higgs couples to $2a$.  Dualizing these phases gives
\begin{equation}
   \calL_{\rm a}
   =\frac{3r}{4\pi}a\,\dd a
   +\frac{1}{2\pi}\beta_b\dd(A-3a)
   +\frac{1}{2\pi}\beta_P\dd(2a).
   \label{eq:BFstartMain}
\end{equation}
The dynamical $K$ matrix has a null vector for every integer $r$.  After an integral change of variables, the massive block has determinant $-1$, while the remaining compact field $\tilde a$ has no Chern-Simons term.  The universal charge-sector response is
\begin{equation}
   \calL_{2e}=\frac{\kappa}{2}(\dd\tilde a)^2+
   \frac{2}{2\pi}\tilde a\dd A+\cdots .
   \label{eq:chargeSC}
\end{equation}
A $2\pi$ monopole of $\tilde a$ creates $b^2\det P$, so Eq.~\eqref{eq:chargeSC} is the dual response of a charge-$2e$ superconductor.  The quotient in Eq.~\eqref{eq:U3quotient} fixes the flux normalization and removes the apparent charge-$6e$ ambiguity of a lifted $SU(3)\times U(1)$ description.  The unimodular reduction is given in the End Matter.

\begin{figure}[tbp]
\centering
\begin{tikzpicture}[x=0.72cm,y=0.84cm]
\tikzset{
  lab/.style={align=center,font=\footnotesize},
  smalllab/.style={align=center,font=\scriptsize},
  axis/.style={font=\footnotesize}
}
\def\figW{4.7}
\def\figHH{2.90}
\def\figMU{1.05}

\begin{scope}
\node[anchor=west,font=\footnotesize] at (-4.85,3.6) {(a) $\mu=0$};

\fill[softorange]
  (-\figW,0)--(-\figW,\figHH)--(-\figHH,\figHH)--(0,0)--cycle;
\fill[softpurple]
  (0,0)--(-\figHH,\figHH)--(\figHH,\figHH)--cycle;
\fill[softgreen]
  (0,0)--(\figHH,\figHH)--(\figW,\figHH)--(\figW,0)--cycle;

\draw[->,line width=0.65pt]
  (-\figW-0.25,0)--(\figW+0.35,0)
  node[right,axis] {$m$};
\draw[->,line width=0.65pt]
  (0,0)--(0,\figHH+0.35)
  node[above,axis] {$|\Delta|$};

\draw[line width=0.95pt] (0,0)--(-\figHH,\figHH);
\draw[line width=0.95pt] (0,0)--(\figHH,\figHH);

\fill[softblue!90!black] (0,0) circle (1.8pt);
\node[
  smalllab,
  anchor=north,
  text=softblue!90!black
] at (0,-0.04) {multicritical point};

\node[lab] at (-3.25,1.20)
  {$SO(3)^s_0$\\strong-pairing SC\\$c_-=3$};

\node[
  lab,
  fill=softpurple,
  fill opacity=0.90,
  text opacity=1,
  inner sep=1.5pt
] at (0,2.05)
  {$SO(3)^s_1$\\topological SC\\$c_-=3/2$};

\node[lab] at (3.27,1.20)
  {$SO(3)^s_2$ SC$^\ast$\\$\mathbb Z_3$ TO\\$c_-=1$};

\node[smalllab,anchor=north] at (-3.05,-0.14)
  {Mott/CDW\\$\sigma_{xy}=0$\\$c_-=0$};

\node[smalllab,anchor=north] at (3.05,-0.14)
  {$\nu=1/3$ FCI\\
   $\sigma_{xy}=\frac{1}{3}\frac{e^2}{h}$\\
   $c_-=1$};

\node[smalllab,anchor=south east]
  at (-\figHH+0.70,\figHH)
  {$m=-|\Delta|$};

\node[smalllab,anchor=south west]
  at (\figHH-0.65,\figHH)
  {$m=|\Delta|$};
\end{scope}

\begin{scope}[yshift=-4.75cm]
\node[anchor=west,font=\footnotesize] at (-4.85,3.6) {(b) $\mu>0$};

\fill[softorange]
  (-\figW,0)--(-\figW,\figHH)
  -- plot[domain=\figHH:0,samples=100]
     ({-sqrt(\figMU*\figMU+\x*\x)},\x)
  -- cycle;

\fill[softpurple]
  plot[domain=0:\figHH,samples=100]
     ({-sqrt(\figMU*\figMU+\x*\x)},\x)
  -- (-\figW,\figHH)
  -- (\figW,\figHH)
  -- plot[domain=\figHH:0,samples=100]
     ({sqrt(\figMU*\figMU+\x*\x)},\x)
  -- cycle;

\fill[softgreen]
  plot[domain=0:\figHH,samples=100]
     ({sqrt(\figMU*\figMU+\x*\x)},\x)
  -- (\figW,\figHH)
  -- (\figW,0)
  -- cycle;

\draw[->,line width=0.65pt]
  (-\figW-0.25,0)--(\figW+0.35,0)
  node[right,axis] {$m$};
\draw[->,line width=0.65pt]
  (0,0)--(0,\figHH+0.35)
  node[above,axis] {$|\Delta|$};

\draw[line width=0.9pt]
  plot[domain=0:\figHH,samples=120]
  ({-sqrt(\figMU*\figMU+\x*\x)},\x);

\draw[line width=0.9pt]
  plot[domain=0:\figHH,samples=120]
  ({sqrt(\figMU*\figMU+\x*\x)},\x);

\draw[line width=1.7pt,softblue]
  (-\figMU,0)--(\figMU,0);

\node[
  below,
  font=\scriptsize,
  text=softblue!90!black
] at (-\figMU,-0.05)
  {$-\mu$};

\node[
  below,
  font=\scriptsize,
  text=softblue!90!black
] at (\figMU,-0.05)
  {$\mu$};

\node[
  anchor=north,
  font=\scriptsize,
  align=center,
  text=softblue!90!black
] at (0,-0.34)
  {chargon\\metal};

\node[lab] at (-3.20,1.10)
  {$SO(3)^s_0$};

\node[
  lab,
  fill=softpurple,
  fill opacity=0.90,
  text opacity=1,
  inner sep=1.5pt
] at (0,1.55)
  {$SO(3)^s_1$};

\node[lab] at (3.20,1.10)
  {$SO(3)^s_2$};

\node[smalllab,anchor=north] at (-3.05,-0.14)
  {Mott/CDW};

\node[smalllab,anchor=north] at (3.05,-0.14)
  {$\nu=1/3$ FCI};
\end{scope}

\end{tikzpicture}

\caption{
Color-isotropic \(P=\Delta\mathbf 1_3\) branch.
The labels \(SO(3)^s_n\) denote the neutral sector, while all
\(P\)-condensed regions share the charge-\(2e\) response in
Eq.~\eqref{eq:chargeSC}.
(a) At \(\mu=0\), the three superconducting branches meet the adjacent
Mott/CDW and FCI states at a multicritical point.
(b) At \(\mu>0\), the same three branches surround a chargon-metal
window.
}
\label{fig:PI}
\end{figure}

\prlsection{Color-isotropic $P=\Delta\mathbf 1_3$ branch}
We first take
\begin{equation}
   \langle P\rangle=\Delta\mathbf 1_3 .
   \label{eq:Pdelta}
\end{equation}
We name pairing chirality by the gauge-invariant electron-pair amplitude $\langle\psi_{{\rm el},\bk}\psi_{{\rm el},-\bk}\rangle\propto b^2\det P(\bk)$, following Refs.~\cite{ReadGreen2000,WangZaletel2025}.  The target branch uses
\begin{equation}
   P(\bk)=\Delta\,(k_x-i k_y)\mathbf 1_3 ,
   \label{eq:PwaveIso}
\end{equation}
so
\begin{equation}
   \det P(\bk)=\Delta^3(k_x-i k_y)^3,
   \label{eq:fif}
\end{equation}
and the physical Cooper pair has $f-if$ chirality.  With this convention, the middle branch of Fig.~\ref{fig:PI} is the desired $c_-=3/2$ superconductor.

The stabilizer of this condensate is $O(3)$ inside $U(3)$.  After determinant locking, the internal neutral stabilizer is
\begin{equation}
   \{g\in U(3):\det g=1,
   \ gg^T=1\}=SO(3).
   \label{eq:SO3higgs}
\end{equation}
Thus every region in this branch is the charge-$2e$ superconductor of Eq.~\eqref{eq:chargeSC}, stacked with an $SO(3)$ neutral sector.  The $U(3)$ quotient fixes the allowed charge-vortex fluxes but leaves the neutral gauge group $SO(3)$.

The neutral response follows from the paired BdG topology.  Choosing a
gauge in which the color-isotropic condensate is real and positive,
\(\langle P\rangle=|\Delta|\mathbf 1_3\), we decompose the complex
\(SO(3)\)-vector fermion into two real vector-Majorana fields,
\begin{equation}
   f_a=\frac{1}{\sqrt2}
   \left(\chi_{+,a}+i\chi_{-,a}\right).
   \label{eq:fMajorana}
\end{equation}
The two Majorana blocks have masses \(m\pm|\Delta|\).  A minimal
regularized Hamiltonian for each block is
\begin{equation}
  h_\pm(\bk)
  =
  k_1\sigma_x+k_2\sigma_y
  +
  \left(m\pm|\Delta|-Bk^2\right)\sigma_z,
  \qquad B>0,
  \label{eq:BdG}
\end{equation}
where the Pauli matrices act in the two-component Dirac space.  Let
\(\nu_\pm\in\{0,1\}\) denote the regulated BdG Chern index of the
\(\chi_\pm\) block.  With the convention in Eq.~\eqref{eq:BdG},
\begin{equation}
  \nu_\pm=\Theta\!\left(m\pm|\Delta|\right),
  \qquad
  n=\nu_++\nu_- .
  \label{eq:nudef}
\end{equation}
Thus \(n=0,1,2\) counts the number of topological
\(SO(3)\)-vector Majorana blocks.  Each topological block induces one
unit of the spin \(SO(3)\) Chern-Simons level, giving
\begin{equation}
   \calL_{\rm neutral}^{(n)}
   =
   3\Om-SO(3)^s_n .
   \label{eq:SO3n}
\end{equation}
For \(SO(3)^s_n\), equivalently the integer-spin sector of
\(SU(2)_{2n}\),
\begin{equation}
   c_-\!\left[SO(3)^s_n\right]
   =
   \frac{3n}{n+1},
   \qquad n=0,1,2,
\end{equation}
so the physical chiral central charge is
\(c_-=3-3n/(n+1)\)~\cite{WittenCS1989,Kitaev2006}:
\begin{center}
\renewcommand{\arraystretch}{1.3}
\begin{ruledtabular}
\begin{tabular}{c c c c}
region & \((\nu_+,\nu_-)\) & neutral sector & \(c_-\) \\
\hline
\(m>|\Delta|\) & \((1,1)\) & \(3\Om-SO(3)^s_2\) & \(1\) \\
\(-|\Delta|<m<|\Delta|\) & \((1,0)\) &
\(3\Om-SO(3)^s_1\) & \(3/2\) \\
\(m<-|\Delta|\) & \((0,0)\) & \(3\Om\) & \(3\) \\
\end{tabular}
\end{ruledtabular}
\end{center}
All three regions share the charge-\(2e\) response in
Eq.~\eqref{eq:chargeSC}.  They differ only in the neutral color
response.

The $SO(3)^s_2$ superconductor is the pure-Higgs descendant of the $C_f=1$ FCI. Condensing $P=\Delta\mathbf 1_3$ in the determinant-fixed $U(3)^s_{1,1}$ state leaves the charge sector in Eq.~\eqref{eq:chargeSC} and restricts the neutral gauge field to the real $SO(3)\subset SU(3)$ subgroup. The fundamental ${\bf 3}_{SU(3)}$ becomes the real vector of $SO(3)$, and the embedding has Dynkin index two~\cite{BaisBouwknegtSurridgeSchoutens1988,LackiZaugg1989},
\begin{equation}
   \left. SU(3)^s_1\right|_{SO(3)}=SO(3)^s_2 .
   \label{eq:SU3SO3embedding}
\end{equation}
Equivalently, this is the $N=3$ case of the small-level relation $SO(N)_2\leftrightarrow SU(N)_1$ used in Ref.~\cite{AharonyBeniniHsinSeiberg2017}. BdG-wise, the complex $C_f=1$ color band becomes two topological real vector-Majorana bands, so $(\nu_+,\nu_-)=(1,1)$.

This branch is an SC$^\ast$: its superconducting sector is the ordinary charge-$2e$ sector of Eq.~\eqref{eq:chargeSC}, while its neutral complement carries the same electronic Laughlin order as the parent FCI. The End Matter uses the spin level-rank duality to obtain
\begin{equation}
3\Om-SO(3)^s_2\simeq U(1)_3 .
\label{eq:SO3LaughlinMain}
\end{equation}
Thus the pure-Higgs branch inherits both the chiral $\mathbb Z_3$ topological order and the chiral central charge $c_-=1$ of the parent $\nu=1/3$ FCI, and need not carry density order.

The $SO(3)^s_1$ superconductor is the new branch.  Exactly one real color-Majorana block remains topological.  The neutral response contains $SO(3)^s_1$, giving three chiral Majorana modes and
\begin{equation}
   c_-=3-\frac{3}{2}=\frac{3}{2} .
   \label{eq:c32}
\end{equation}
This phase has the pairing symmetry and topology found numerically~\cite{WangZaletel2025,GuerciAbouelkomsanFu2026VortexLattice}.  It appears because the color-isotropic pair Higgs resolves the complex color band into two real vector-Majorana blocks.

The $SO(3)^s_0$ branch is the strong-pairing superconductor with $c_-=3$.  Both real color-Majorana blocks are inverted, no neutral Chern-Simons sector remains, and the response is only the fixed invertible background $3\Om$.  Equivalently, $(\nu_+,\nu_-)=(0,0)$ is adiabatically connected to the BEC limit of the paired anyon fluid.  The remaining low-energy charged objects are $P\sim ff$, each carrying charge $2e/3$ in the determinant-Higgs background~\cite{WangZaletel2025,ShiSenthil2025PRX,ShiSenthil2025NonAbelian,FanVishwanathWang2026}.

\prlsection{Finite density and criticality}
The same determinant-locked charge sector also describes finite density.  Doping the electron density by $\delta$ creates total $f$ density $n_f=3|\delta|$.  A color-neutral saddle has
\begin{equation}
   \langle f^\dagger_a f_b\rangle=\frac{n_f}{3}\delta_{ab},
   \qquad
   \langle f^\dagger T^A f\rangle=0,
   \label{eq:colorsinglet}
\end{equation}
for every traceless $SU(3)$ generator $T^A$.  The non-Abelian Gauss law then gives zero average color flux,
\begin{equation}
   \frac{k}{2\pi}\langle F^A_{\alpha,12}\rangle
   +\langle f^\dagger T^A f\rangle=0
   \quad\Rightarrow\quad
   \langle F^A_{\alpha,12}\rangle=0 .
   \label{eq:Gauss}
\end{equation}
The doped state is therefore a chargon metal of charge-$e/3$ partons.  The saddle carries no net color charge, while the parton Fermi surface consists of three fundamental gauge-color components with total density $3|\delta|$.  The state is compressible, has zero average Cartan color flux, and has a finite clean-limit Drude weight~\cite{ZhangHolonMetal2025}.

When the chemical potential enters the parton band, the BdG boundaries deform to
\begin{equation}
   m=\pm R_\mu,
   \qquad R_\mu=\sqrt{|\Delta|^2+\mu^2}.
   \label{eq:Rmu}
\end{equation}
At finite density, $|m|<|\mu|$, and hence weak pairing lies on the middle BdG branch.  The Fermi surface then pairs into the $SO(3)^s_1$ phase with $c_-=3/2$.

The zero-density limit is more subtle.  The direct FCI-to-CDW route involves a parton band inversion and hence a Dirac gap closing.  The superconducting phase emanates from the associated multicritical point along the pair-Higgs direction.  At that point the gauge-invariant electron channel is gapless because all three parton constituents are gapless.

\prlsection{Translation-symmetric construction}
A second construction keeps all three magnetic valleys and preserves microscopic translations by condensing the full-rank color-valley pair field
\begin{equation}
   \Phi^a_I
   =\frac{1}{2}\epsilon^{abc}\epsilon_{IJK}
   \left\langle f_{bJ}D_-f_{cK}\right\rangle .
   \label{eq:CVLphi}
\end{equation}
A full-rank condensate $\langle\Phi^a_I\rangle=\Delta\delta^a_I$ locks magnetic valley to gauge color.  Its stabilizer inside $SU(3)_g$ is trivial, so it completely Higgses the color gauge group while preserving microscopic translations.  This is analogous to color-flavor locking in dense QCD~\cite{AlfordRajagopalWilczek1999}.

After locking, all nine fields remain and resolve into one $\bk=0$ sector and four pairs at opposite ordinary crystal momenta.  When all nine sectors are in the strong-pairing regime, only the fixed $3\Om$ response remains.  This gives a translation-preserving realization of the $c_-=3$ $SO(3)^s_0$ molecular superconductor discussed above.

A translation-symmetric dispersion can instead place the $\bk=0$ sector and one pair at $\pm\bk_0$ in the $C_{\rm BdG}=1$ regime, while the other three momentum pairs remain topologically trivial.  The resulting phase has three chiral Majorana modes and $c_-=3/2$, giving a translation-preserving realization of the $SO(3)^s_1$ topological superconductor discussed above.  Because the full-rank lock removes the neutral gauge sector, it cannot realize the $SO(3)^s_2$ SC$^\ast$ branch with residual $U(1)_3$ topological order.

For $\langle\Phi^a_I\rangle=\Delta\delta^a_I$, the diagonal color-valley fields $f_{11},f_{22},f_{33}$ span the $\bk=0,\pm\bk_0$ sectors.  Restricting the color gauge field to its Cartan sector gives
\begin{equation}
K=\begin{pmatrix}2&1\\1&2\end{pmatrix},\label{eq:threePocketKMain}
\end{equation}
and $q_1=(1,0)$, $q_2=(0,1)$, and $q_3=(-1,-1)$.  This reproduces the three-pocket $U(1)^2$ theory of Ref.~\cite{FanVishwanathWang2026}.  The three inter-pocket pair condensates span the full Cartan charge lattice and completely Higgs both $U(1)$ gauge fields.  After particle-hole conjugation, the strong- and weak-pairing phases of Ref.~\cite{FanVishwanathWang2026} give the $c_-=3$ and $c_-=3/2$ branches above.  The magnetic translations and Cartan reduction are detailed in the End Matter.

\prlsection{Discussion}
We have presented a determinant-locked $U(3)$ parton route from a $\nu=1/3$ FCI to charge-$2e$ superconductivity.  The color-isotropic branch gives an SC$^\ast$ with $c_-=1$, the target $SO(3)^s_1$ superconductor with $c_-=3/2$, and a strong-pairing molecular superconductor with $c_-=3$.  The same infrared theory contains the nearby $\sigma_{xy}=0$ CDW/Mott state and the finite-density chargon metal.

At zero doping, the superconducting phase emanates from the FCI-to-CDW multicritical point.  The electron gap softens there.  In the translation-broken construction, the momentum of $b^2\det P$ depends on which valley sector is retained for each color.  We focus on the zero-momentum choice represented by the color-isotropic theory, while other choices can give a pair-density wave.  Numerics can also search for the nearby $c_-=1$ SC$^\ast$ branch and the $c_-=3$ strong-pairing branch using entanglement spectra, edge central charge, and modular data~\cite{LiHaldane2008,ZaletelMongPollmann2013,NiuThoulessWu1985}.

A full-rank color-valley lock gives translation-preserving realizations of the $c_-=3$ molecular and $c_-=3/2$ topological superconductors corresponding to the $SO(3)^s_0$ and $SO(3)^s_1$ color-isotropic branches.  Its three-mode Cartan description is the particle-hole conjugate of Ref.~\cite{FanVishwanathWang2026}, while the full condensate removes the entire neutral gauge sector.  Existing experiments do not yet determine whether the superconducting phase preserves microscopic translations.  Since both realizations have the ordinary charge-$2e$ sector of Eq.~\eqref{eq:chargeSC}, Little-Parks and Josephson measurements should see the conventional $hc/2e$ period.  Thermal Hall transport measures $c_-$, while edge tunneling probes the chiral Majorana boundary modes.

\textbf{Acknowledgments.}
We especially thank Ashvin Vishwanath and Chong Wang for discussions on magnetic translations.  We also thank Da-Chuan Lu, Ya-Hui Zhang, Zhaoyu Han, Zhi-Qiang Gao, and Zhengyan Darius Shi for helpful discussions.  T.W. is grateful for support from the Harvard Quantum Initiative Fellowship and the Simons Collaboration on Ultra-Quantum Matter, a grant from the Simons Foundation (Grant No. 651440).

\bibliographystyle{apsrev4-2}
\bibliography{main}

\begin{center}
\textbf{End Matter}
\end{center}

\prlsection{Magnetic translations}
Let
\begin{equation}
X=\begin{pmatrix}0&1&0\\0&0&1\\1&0&0\end{pmatrix},
\quad
Z=\begin{pmatrix}1&0&0\\0&\omega&0\\0&0&\omega^2\end{pmatrix},
\quad
XZ=\omega ZX .
\label{eq:clockshiftEnd}
\end{equation}
where $\omega=e^{2\pi i/3}$.
In one gauge, translations are accompanied by color rotations,
\begin{equation}
\begin{aligned}
   T_1:f_a(x,y)&\mapsto X_{ab}f_b(x+1,y),
   \\
   T_2:f_a(x,y)&\mapsto Z_{ab}f_b(x,y+1).
\end{aligned}
   \label{eq:colorRotationTranslationsEnd}
\end{equation}
Their commutator is the center element $\omega\mathbf 1_3$, which acts trivially on gauge-singlet electron operators.  The site-dependent gauge transformation $f_a(x,y)\mapsto (Z^yX^x)_{ab}f_b(x,y)$ gives an equivalent gauge,
\begin{equation}
\begin{aligned}
   T_1:f_a(x,y)&\mapsto f_a(x+1,y),
   \\
   T_2:f_a(x,y)&\mapsto\omega^x f_a(x,y+1).
\end{aligned}
   \label{eq:LandauGaugeEnd}
\end{equation}
Equation~\eqref{eq:LandauGaugeEnd} makes the counting explicit: every color has a three-site magnetic unit cell and hence three folded pockets.  Thus both gauges contain the nine fields $f_{aI}$, with $a,I=1,2,3$, rather than only three.  The color rotations in Eq.~\eqref{eq:colorRotationTranslationsEnd} alone are gauge removable and do not constitute color-valley locking.  Locking requires the full-rank condensate $\langle\Phi^a_I\rangle=\Delta\delta^a_I$, which completely Higgses $SU(3)_g$. After the gauge transformation above, the same condensate has a site-dependent color orientation.

\prlsection{Color-valley locking and Cartan reduction}
With translations unbroken, organize the nine reduced-zone fields as the matrix $\Psi_{aI}$, with color on the left and magnetic valley on the right.  For the condensate $\langle\Phi^a_I\rangle=\Delta\delta^a_I$, microscopic translations preserve it by combining the valley action with a compensating color gauge rotation:
\begin{equation}
\begin{aligned}
   T_1:\Psi(\br)&\mapsto X\Psi(\br+\ell_1)X^\dagger,\\
   T_2:\Psi(\br)&\mapsto Z^\dagger\Psi(\br+\ell_2)Z .
\end{aligned}
   \label{eq:lockedTranslationsEnd}
\end{equation}
The left matrices act on color and the right matrices act on valley.  Their opposite center phases cancel, so the two translations commute on the locked fields.  The nine matrices $X^nZ^m$ diagonalize Eq.~\eqref{eq:lockedTranslationsEnd} and carry ordinary crystal momenta
\begin{equation}
   \bk_{mn}=\frac{m\mathbf G_1+n\mathbf G_2}{3}
   \quad (m,n\in\mathbb Z_3).
   \label{eq:kgridEnd}
\end{equation}
The $3\times3$ momentum grid contains one self-conjugate sector at $\bk=0$ and four pairs $(\bk,-\bk)$.  The diagonal subspace $\psi_i=f_{ii}$ is spanned by $\mathbf 1_3,Z,Z^2$, whose Fourier combinations carry $\bk=0,\pm\bk_0$.  The six off-diagonal fields form the other three momentum pairs.  Once the full-rank lock Higgses $SU(3)_g$, a translation-symmetric dispersion can place the diagonal three sectors in the $C_{\rm BdG}=1$ regime while keeping the other six topologically trivial.

To expose the relation to Ref.~\cite{FanVishwanathWang2026}, write the color gauge field in the Cartan basis,
\begin{equation}
   \alpha=\operatorname{diag}(a_1,a_2,-a_1-a_2).
   \label{eq:CartanGaugeEnd}
\end{equation}
Substitution into the level-one color Chern-Simons term gives Eq.~\eqref{eq:threePocketKMain}, while $\psi_i$ carry the three charge vectors listed there.  Since $q_i^{\sf T}K^{-1}q_j=\delta_{ij}-1/3$, this is the three-pocket flux-attachment theory of Ref.~\cite{FanVishwanathWang2026}.

Within the diagonal three-field sector, the locked pair components reduce to
\begin{equation}
\begin{aligned}
\left.\Phi^1_1\right|_{\rm diag}&\sim\psi_2D_-\psi_3,\\
\left.\Phi^2_2\right|_{\rm diag}&\sim\psi_3D_-\psi_1,\\
\left.\Phi^3_3\right|_{\rm diag}&\sim\psi_1D_-\psi_2.
\end{aligned}
\label{eq:FanPairsEnd}
\end{equation}
Their Cartan charges are $(-1,0)$, $(0,-1)$, and $(1,1)$, which span the full integer charge lattice.  They Higgs both Cartan gauge fields, while the same full-rank condensate Higgses the six off-diagonal $SU(3)$ gauge bosons.  Thus no neutral gauge sector remains.  The product $b^2\det\Phi\sim b^2\Delta_{12}\Delta_{23}\Delta_{31}$ is the charge-$2e$ superconducting order parameter of Ref.~\cite{FanVishwanathWang2026}.

Ref.~\cite{FanVishwanathWang2026} uses the $\nu=2/3$ convention, where the strong- and weak-pairing phases have $c_-=-2$ and $c_-=-1/2$.  Particle-hole conjugation sends $c_-\mapsto1-c_-$, giving $c_-=3$ and $c_-=3/2$ in the $\nu=1/3$ convention used here.

\prlsection{Charge-sector derivation}
The trace-sector reduction in Eq.~\eqref{eq:chargeSC} follows only from the two Higgs charges.  In a local $SU(3)\times U(1)$ lift, $b$ has physical charge $+1$ and trace gauge charge $-3$, while the common phase of either full-rank pair field, $P$ or $\Phi$, has trace gauge charge $+2$.  Dualizing the condensed phases gives
\begin{equation}
   \calL_{\rm tr}
   =\frac{3r}{4\pi}a\,\dd a
   +\frac{1}{2\pi}\beta_b\dd(A-3a)
   +\frac{1}{2\pi}\beta_P\dd(2a).
   \label{eq:BFstart}
\end{equation}
With $A$ suppressed, the dynamical matrix in the basis $(a,\beta_b,\beta_P)$ is
\begin{equation}
K_{\rm full}=\begin{pmatrix}
3r&-3&2\\
-3&0&0\\
2&0&0
\end{pmatrix},
\qquad
\det K_{\rm full}=0 .
\label{eq:Kfull}
\end{equation}
The integral change of variables
\begin{equation}
   \lambda=-3\beta_b+2\beta_P,
   \qquad
   \tilde a=-\beta_b+\beta_P
   \label{eq:unimod}
\end{equation}
gives
\begin{equation}
   \calL_{\rm tr}
   =\frac{3r}{4\pi}a\,\dd a
   +\frac{1}{2\pi}\lambda\dd(a-A)
   +\frac{2}{2\pi}\tilde a\dd A .
   \label{eq:BFend}
\end{equation}
The $(a,\lambda)$ block has
\begin{equation}
   K_{\rm massive}=\begin{pmatrix}3r&1\\1&0\end{pmatrix},
   \qquad
   \det K_{\rm massive}=-1,
   \label{eq:Kmassive}
\end{equation}
and leaves no Abelian topological order.  The remaining compact field $\tilde a$ has no Chern-Simons term, and its $2\pi$ monopole creates $b^2\det P$ or $b^2\det\Phi$.  The genuine $U(3)$ quotient fixes the flux normalization and removes the apparent charge-$6e$ ambiguity of the lifted description.

\prlsection{Level-rank duality and Laughlin dictionary}
We use the superscript $s$ for spin Chern-Simons gauge fields.  The notation $X-Y$ means stacking $X$ with the orientation reverse of $Y$.  The orthogonal spin level-rank duality is
\begin{equation}
   SO(N)^s_k\times SO(0)^s_1
   \simeq
   SO(k)_{-N}\times SO(Nk)^s_1 .
   \label{eq:SOspinLevelRank}
\end{equation}
The factor $SO(0)^s_1$ keeps the transparent local fermion explicit, while $SO(Nk)^s_1$ is an invertible spin TQFT with framing anomaly $c=Nk/2$~\cite{AharonyBeniniHsinSeiberg2017}.  For $N=3$ and $k=2$,
\begin{equation}
\begin{aligned}
   SO(3)^s_2\times SO(0)^s_1
   &\simeq SO(2)_{-3}\times SO(6)^s_1\\
   &\simeq U(1)_{-3}\times SO(6)^s_1 .
\end{aligned}
   \label{eq:SO3LRend}
\end{equation}
The invertible factor $SO(6)^s_1$ has $c_-=3$ and represents $3\Om$, giving
\begin{equation}
   3\Om-SO(3)^s_2\simeq U(1)_3 .
   \label{eq:SO3toU13End}
\end{equation}
The parent FCI has the equivalent presentation
\begin{equation}
   3\Om-SU(3)^s_1\simeq U(1)_3 .
   \label{eq:SU3toU13End}
\end{equation}

The $SU(3)_1$ theory has the Abelian lines $1,{\bf3},\bar{\bf3}$.  A useful cover of $SO(3)^s_2$ is $\mathrm{Spin}(3)_2=SU(2)_4$, whose integer-spin sector contains $j=0,1,2$.  In the Higgs completion the bosonic $j=2$ simple current is screened, and the $j=1$ fixed point resolves into two Abelian lines $x,x^2$ with
\begin{equation}
   x^3=1,
   \qquad
   x\simeq{\bf3},
   \qquad
   x^2\simeq\bar{\bf3},
   \qquad
   \theta_x=\theta_{x^2}=e^{-2\pi i/3} .
\end{equation}
In the Laughlin presentation, let $q^3=\psiel$ and $\theta_q=e^{\pi i/3}$.  Then
\begin{equation}
   x\longleftrightarrow q\psiel,
   \qquad
   x^2\longleftrightarrow q^2 .
\end{equation}
The equivalence is one of electronic spin topological orders, with local-electron attachment understood, rather than an equality of bosonic modular tensor categories~\cite{LackiZaugg1989,BaisSlingerland2009}.

\end{document}